\documentclass[smallextended]{svjour3}       
\smartqed  									 

\usepackage{graphicx}
\usepackage{amssymb}   
\usepackage{textcmds}  
\usepackage{color}     
\usepackage{gensymb}   
\usepackage[]{hyperref}

\usepackage{natbib}
\bibpunct{(}{)}{;}{a}{}{,}
\newcommand{\darwin}{D{\sc arwin}}
\newcommand{\darwins}{D{\sc arwin\,}}

\newcommand*\aap{A\&A}

\newcommand*\aj{AJ}
\newcommand*\ao{Appl Opt}

\newcommand*\apj{ApJ}

\newcommand*\apss{Ap\&SS}
\newcommand*\araa{ARA\&A}

\newcommand*\icarus{Icarus}

\newcommand*\nat{Nature}

\newcommand*\pasp{PASP}

\newcommand*\procspie{Proc SPIE}

\begin{document}

\title{Space-based infrared interferometry to study exoplanetary atmospheres}

\titlerunning{Space-based infrared interferometry to study exoplanetary atmospheres} 

\author{D.~Defr\`ere$^1$, A.~L\'eger$^2$, O.~Absil$^1$, C.~Beichman$^3$, B.~Biller$^4$, W.C.~Danchi$^5$, K.~Ergenzinger$^6$, C.~Eiroa$^7$, S.~Ertel$^8$, M.~Fridlund$^{9,16}$, A.~Garc\'ia Mu\~noz$^{10}$, M.~Gillon$^1$, A.~Glasse$^{11}$, M.~Godolt$^{10}$, J.L.~Grenfell$^{12}$, S.~Kraus$^{13}$, L.~Labadie$^{14}$, S.~Lacour$^{15}$, R.~Liseau$^{16}$, G.~Martin$^{17}$, B.~Mennesson$^{18}$, G.~Micela$^{19}$, S.~Minardi$^{20}$, 
S.P.~Quanz$^{21}$, H.~Rauer$^{10,12}$, S.~Rinehart$^5$, N.C.~Santos$^{22,23}$, F.~Selsis$^{24}$, J.~Surdej$^1$, F.~Tian$^{25}$, E.~Villaver$^7$, P.J.~Wheatley$^{26}$, M.~Wyatt$^{27}$.}

\authorrunning{Defr\`ere et al.}

\institute{D.\, Defr\`ere \at
           Tel.: +32-4-3669758\\
           \email{ddefrere@uliege.be}\\           
           \and
           $^1$ Space sciences, Technologies \& Astrophysics Research (STAR) Institute, University of Li\`ege, Li\`ege, Belgium\\
           \and
          $^2$ Institut d'Astrophysique Spatiale, Universit\'e de Paris-Sud, Orsay, France\\
           \and
          $^3$ NASA Exoplanet Science Institute, California Institute of Technology, Pasadena, California, USA\\
           \and
          $^4$ Institute for Astronomy, University of Edinburgh, Edinburgh, United Kingdom\\
           \and
          $^5$ NASA Goddard Space Flight Center, Exoplanets \& Stellar Astrophysics Laboratory, Greenbelt, USA\\
           \and
          $^6$ Airbus Defence and Space GmbH\\
           \and
          $^7$ Dpto. Fisica Teorica, Universidad Autonoma de Madrid, Canto-blanco, 28049 Madrid, Spain\\
           \and
          $^8$ Steward Observatory, Department of Astronomy, University of Arizona, Tucson, Arizona, USA\\
           \and
          $^9$ Leiden Observatory, University of Leiden, PO Box 9513, NL-2300 RA, Leiden, the Netherlands\\
           \and
          $^{10}$ Technische Universit\"at Berlin, Berlin, Germany\\
           \and
          $^{11}$ Astronomy Technology Centre, United Kingdom \\
           \and
          $^{12}$ Institute for Planetary Research, German Aerospace Center, Berlin, Germany\\
           \and
          $^{13}$ School of Physics and Astronomy, University of Exeter, Exeter, United Kingdom\\ 
           \and
          $^{14}$ University of Cologne, Germany\\
           \and
          $^{15}$ LESIA, Observatoire de Paris, PSL Research University, 92195 Meudon Cedex, France\\
           \and
          $^{16}$ Department of Space, Earth and Environment, Chalmers University of Technology, Onsala Space Observatory, 439 92 Onsala, Sweden\\
           \and
          $^{17}$ Institut de plan\'etologie et d'astrophysique de Grenoble, Grenoble, France\\
           \and
          $^{18}$ Jet Propulsion Laboratory, USA\\
           \and
          $^{19}$ Osservatorio Astronomico di Palermo, INAF, Italy\\
           \and
          $^{20}$ innoFSPEC, Leibniz-Institut f\"ur Astrophysik Potsdam (AIP) Germany\\
           \and
          $^{21}$ ETH Zurich, Institute for Particle Physics and Astrophysics, Zurich, Switzerland\\
           \and
          $^{22}$ Instituto de Astrof\'isica e Ci\^encias do Espa\c{c}o, Universidade do Porto, CAUP, Rua das Estrelas, 4150-762 Porto, Portugal\\
           \and
          $^{23}$ Departamento de F\'isica e Astronomia, Faculdade de Ci\^encias, Universidade do Porto, Rua do Campo Alegre, 4169-007 Porto, Portugal\\
           \and
          $^{24}$ University of Bordeaux, France \\
           \and
          $^{25}$ Tsinghua University of Bejing, China\\
           \and
          $^{26}$ Department of Physics, University of Warwick, Gibbet Hill Road, Coventry CV4 7AL, United Kingdom\\
           \and
          $^{27}$ Institute of Astronomy, Madingley Road, Cambridge CB3 0HA, United Kingdom\\
}

\date{Received: date / Accepted: date}
\maketitle

\begin{abstract}
The quest for other habitable worlds and the search for life among them are major goals of modern astronomy. One way to make progress towards these goals is to obtain high-quality spectra of a large number of exoplanets over a broad range of wavelengths. While concepts currently investigated in the United States are focused on visible/NIR wavelengths, where the planets are probed in reflected light, a compelling alternative to characterize planetary atmospheres is the mid-infrared waveband (5-20~$\mu$m). Indeed, mid-infrared observations provide key information on the presence of an atmosphere, the surface conditions (e.g., temperature, pressure, habitability), and the atmospheric composition in important species such as H$_2$O, CO$_2$, O$_3$, CH$_4$, and N$_2$O. This information is essential to investigate the potential habitability of exoplanets and to make progress towards the search for life in the Universe. Obtaining high-quality mid-infrared spectra of exoplanets from the ground is however extremely challenging due to the overwhelming brightness and turbulence of the Earth's atmosphere. In this paper, we present a concept of space-based mid-infrared interferometer that can tackle this observing challenge and discuss the main technological developments required to launch such a sophisticated instrument. 
\end{abstract}

\section{Introduction}

The goal of finding habitable planets and even planets with signs of (primitive) life around other stars is extremely challenging and requires a variety of complementary observations. Experience gained so far in exoplanet atmospheric research shows that a broad wavelength coverage and sufficiently high spectral resolution are required to break the degeneracies in composition and climate associated with retrieval of atmospheric spectra. In that regard, an important wavelength regime is the mid-infrared (5-20~$\mu$m). It provides data to measure key planetary parameters, such as the size, temperature, the presence of an atmosphere, as well as the presence of important atmospheric molecules such as H$_2$O, CO$_2$, O$_3$, CH$_4$, and N$_2$O (see Table~\ref{tab1}). From an observational standpoint, it also has the considerable advantage of offering a more favorable planet/star contrast than at visible wavelengths (10$^{-7}$ vs 10$^{-10}$) for an Earth-Sun system.\\

\begin{table}[!t]
\caption{Information on planets that can be obtained from low-resolution (R$\simeq$20) mid-infrared observations (5-20~$\mu$m). Two continuum (Cont.) bands (2 and 3) are also given, where in a cloud-free atmosphere, emission from the surface might be seen \citep[data from][]{Desmarais:2002,Seager:2016,Airapetian:2017}. Values given for N$_2$O and NO are approximative. {\bf The wavelength columns ($\lambda_{\rm MIN}$, $\lambda_{\rm MAX}$, and $\lambda_{\rm AVG}$) are given in microns. R gives the corresponding spectral resolution.}}
\begin{center}
\begin{tabular}{l l c c c c c c}
\hline
\hline
            		   & 
Information on planet & 
Species  & 
$\lambda_{\rm MIN}$ & 
$\lambda_{\rm MAX}$ & 
$\lambda_{\rm AVG}$ & 
R\\

\hline
1 & Orbit characteristics       & Cont.~1 & 6.00 & 20.0 & 13.0 & 1\\
2 &  Combination of temperature, & Cont.~2 & 10.1 & 12.4 & 11.2 & 5\\
  & radius, and albedo          & Cont.~3 & 8.16 & 9.24 & 8.67 & 8\\
3 &  Existence of atmosphere & Cont.~1 & 6.00 & 20.0 & 13.0 & 1\\
  &                     & CO$_2$ & 9.07 & 9.56 & 9.31 & 19\\
  &						&        & 10.1 & 10.7 & 10.4 & 16\\
  &						&        & 13.3 & 17.0 & 15.0 & 4\\
  &						& NO     & 5.1  & 5.5  & 5.3 & 20\\
4 & Presence of water & H$_2$O & 6.67 & 7.37 & 7.00 & 10\\
  &                   &        & 17.4 & 25.0 & 20.5 & 3  \\
5 & Suggestion of life  & CH$_4$ & 7.37 & 7.96 & 7.65 & 13\\
  &						&        & 7.37 & 8.70 & 7.98 & 6\\
  &						& N$_2$O & 7.50 & 9.00 & 7.25 & 10\\
  &					    & O$_3$  & 9.37 & 9.95 & 9.65 & 17\\
\hline
\end{tabular}
\end{center}
\label{tab1}
\end{table}

Mid-infrared observations from the ground are however very challenging due to the turbulence and brightness of the atmosphere. For instance, the atmosphere is approximately 10 billion times brighter than a 300K Earth-sized planet located at 10\,pc when observed with an 8-m telescope. Overcoming the background photon noise limit alone requires prohibitively long integration times to study rocky exoplanets, even around nearby stars. In addition, because the Earth's atmosphere is mostly opaque at the wavelengths corresponding to major molecular absorption features (such as H$_2$O and CO$_2$), searching for the broad spectral signatures of major molecular species in planetary atmospheres will generally be very difficult from the ground. To overcome these issues, access to space is mandatory. In the short term, the study of transiting systems is highly promising for planets larger than a few Earth radii and will be a major focus for the James Webb Space Telescope \citep{Beichman:2014}. Nevertheless, the rarity of nearby planetary systems suitably aligned to produce a transit, the short duration of the transit, and the noise produced by the host star means that the study of the atmospheres of terrestrial planets, especially those orbiting Sun-like stars, is probably unachievable via this technique. Thus, techniques which separate the light of the planet from the glare of the host star and can characterize exoplanets with any orbital configuration are essential.\\

In order to spatially resolve the systems closest to Earth (within 10\,pc) in the mid-infrared, an aperture of at least 80\,m in diameter would however be required and this is presently not feasible. A way to increase the spatial resolution is to use an interferometer as initially proposed by \citet{Bracewell:1978} and significantly improved by \cite{Angel:1986}. The concept of such an instrument was extensively studied in the 1990s and 2000s by both ESA and NASA. In Europe, ESA focused mainly on the \darwins project \citep[e.g.,][]{Leger:1996,Cockell:2009}, which consisted of a space-based flotilla of mid-infrared telescopes using nulling interferometry as the measurement principle. In the United states, a similar concept, called the Terrestrial Planet Finder Interferometer \citep[TPF-I,][]{Angel:1997}, was considered as the final piece of NASA's ambitious Navigator program to characterize Earth-like exoplanets \citep{Lawson:2006}. Between 1996 and 2007, considerable efforts have been carried out by both agencies to define the best mission design and to advance the technologies required for such an ambitious endeavor. Several mission architectures were proposed, key enabling technologies were developed and demonstrated on laboratory test-benches, and advanced data reduction techniques were investigated. These efforts, which resulted in hundreds of articles in the technical literature, culminated in 2007 with the convergence and consensus on a mission architecture called the Emma X-array {\bf \citep{Cockell:2009}}. In parallel, both agencies also appointed teams to investigate the scientific issues related to the search for life on exoplanets \citep[e.g.,][]{Lawson:2007,Fridlund:2010}. Some of the key questions are: What are the atmospheric compositions of rocky exoplanets? Are they habitable? What is a biosignature? Do exoplanets show signs of biological activity? How common is a planet like Earth? How do rocky planets form? While most \darwins and TPF-I activities stopped after 2007 because of fundings reasons, these scientific questions are still a central focus. Today, the exoplanet landscape has greatly changed compared to 2007 but the general consensus in the exoplanet community is still the same: mid-infrared spectra will be required to tackle these fundamental questions {\bf \citep[see e.g.,][]{Fujii:2018}}.\\

\section{Studying planetary atmospheres in the mid-infrared} \label{sec:science}

\subsection{Presence of an atmosphere and basic planetary properties}\label{sec:phasecurves}

For any atmospheric composition, monitoring the variations of thermal emission of an exoplanet during its orbital motion provides a fundamental constraint on meridional transport (hence atmospheric mass) as well as on cloud coverage \citep[e.g.,][]{Selsis2004}. With an increasing number of thermal emission bands monitored, additional planetary properties can be inferred or constrained by orbital photometry: rotation, albedo, obliquity, radius \citep{Selsis2011,Cowan2012,Maurin2012,Selsis2013}, presence of a large satellite \citep{Moskovitz2009}, response to eccentricity \citep{Bolmont2016}. Thermal phase curves have successfully been used to characterize unresolved transiting \citep{Knutson2007,Stevenson:2014} and even non-transiting giant planets \citep{Crossfield2010} but the stellar variability and the required photometric precision make these measurements very difficult. With directly imaged planets, this method will show its full potential. Although the inner working angle will limit the access to the smallest phase angles, only a moderate photometric precision of $\sim 10$\% will be required to achieve a crucial diagnostic and first classification of the targets. Another strength of the method comes from the favorable distribution of orbit orientations. For the median inclination of randomly oriented orbits ($60^{\circ}$), the amplitude of phase curves is only decreased by 10~\% compared with the maximum variations, reached for a  $90^{\circ}$ inclination \citep{Maurin2012}. Only a few observations covering one orbit would be sufficient to start deriving constraints on the climate and several low-resolution or broadband observations would be necessary to resolve the orbital motion of the planet. {\bf Note finally that a by-product of thermal emission monitoring is the orbit characteristics, such as semi-major axis, period, inclination, and position angle.} 

\subsection{Surface conditions, atmospheric composition, and habitability}

In addition to the constraints from orbital broadband photometry, mid-infrared spectroscopy is fundamental to investigate the nature of planetary atmospheres by detecting spectral features, constraining the temperature and pressure structure, the cloudiness and by determining whether the planet could potentially be habitable. Indeed, several important species have mid-infrared spectral signatures that can be detected at low to medium spectral resolving power (e.g., H$_2$O, CO$_2$, O$_3$, CH$_4$, N$_2$O). {\bf The wavelength (minimum, maximum, and average) and corresponding spectral resolution for each spectral feature are given in Table~\ref{tab1}}). In addition, surface conditions (temperature and pressure) can be characterised relatively well from mid-infrared observations (to within $\sim$10 K at 3-$\sigma$) with S/Ns between 10 and 30, depending on spectral resolution \citep{vonParis:2013}. By observing a large number of Habitable Zone (HZ) rocky exoplanets, it will be possible to correlate the concept of habitability with key parameters and processes like spectral type of the parent star, degree of stellar activity, the temperature/pressure structure of the atmosphere, gaseous composition, the circulation and heat transfer of the atmosphere, the {atmospheric chemistry} and photochemistry, the outgassing of atmospheric species, and the presence of a magnetic field.


\subsection{Understanding the concept of biosignature}

A biosignature can be defined as ``an observable feature of a planet, such as its atmospheric composition, that our present models cannot reproduce when including the abiotic physical and chemical processes we know about" \citep{Leger:2011}. {\bf In a recent review, \cite{Catling:2018} define a biosignature as ``any phenomenon, substance, or group of substances that provides evidence of the presence of life.'' These include biosignature gases in the atmosphere, oxygen (O$_2$), ozone (O$_3$), methane (CH$_4$), nitrous oxide (N$_2$O), and methyl chloride (CH$_3$Cl); and a surface biosignature, the vegetation ``red edge,'' a unique reflectance spectrum from plant leaves and a sign of oxygenic photosynthesis. Over the past few years, researchers have probed the efficacy of these signs of biogenic gases and photosynthesis for exoplanets and identified additional candidates. In particular, \cite{Schwieterman:2018} have presented an exhaustive review of the mechanisms, sources, sinks, and environmental by-products for gas and surface biosignature candidates to date.

While the scientific community keeps improving the list of potential biosignature species \citep[e.g.,][]{Seager:2016,Kiang:2018},} it appears clear that our theoretical models are currently limited and the emerging vision is that studying the atmospheres of a large number of exoplanets will be required to provide an essential context for interpreting possible detections of bio-signatures. A good approach could be to create the planetary equivalent of a ``Hertzsprung-Russel`` diagram including species suggestive of life. Only after observing a sufficient number of planetary atmospheres, will it be possible to identify possible anomalies in this diagram that cannot be explained without the presence of life. To make progress in that direction, the mid-infrared regime has a key role to play since it contains spectral signatures of important molecules such as H$_2$O, CO$_2$, O$_3$, N$_2$O, and CH$_4$. In fact, the triple signature (i.e., O$_3$, CO$_2$, H$_2$O) was considered as the most robust indicator for life at the time of the \darwin/TPF studies and would still today be considered as a serious hint of biological activity. The advantage of O$_3$ over O$_2$ is that O$_3$ is a highly sensitive indicator for the existence of even a trace amount of O$_2$ and hence easier to detect at low O$_2$ concentrations \citep{Desmarais:2002}. It is also difficult to produce abiotic O$_3$ if water is present, due to catalytic cycles initiated by water photolysis, which removes O$_3$. {\bf An in-depth review of O$_2$ as a biosignature, rigorously examining the nuances of false positives and false negatives for evidence of life can be found in \cite{Meadows:2018}.}

\section{Space-based interferometer concept} \label{sec:tech}
\subsection{Extracting the planetary photons by nulling interferometry}\label{sec:nulling}

The basic principle of nulling interferometry is to combine the beams coming from two telescopes with a 180 degree phase shift so that a dark fringe appears on the line of sight towards the hosts star, which strongly suppresses its light. On the other hand, off-axis emission, such as that of a planet, can be transmitted by optimizing the baseline length since the nearest bright fringe is located $\lambda$/2/baseline from the dark fringe. However, even when the stellar emission is sufficiently reduced, it is generally not possible to detect Earth-like planets with a static array configuration, because their emission is dominated by the thermal contribution of a series of other extraneous and generally dominant signals originating from the telescope itself (thermal background, readout noise), material in the Solar system (the local zodiacal emission), or around the target star (exozodiacal light). This is the reason why Bracewell proposed to rotate the interferometer so that the planetary signal gets modulated by alternatively crossing high and low transmission regions, while the stellar signal and the background emission remain constant. The planetary signal can then be retrieved by synchronous demodulation. This modulation technique is in many ways similar to the use of a chopper wheel that allows the detection of infrared sources against a thermal background and/or drifting detector offsets. 


\subsection{The Emma X-array configuration}\label{sec:emma}

During the \darwin/TPF-I studies, it was quickly realized that the interferometric array cannot be rotated sufficiently fast to mitigate low frequency instrumental drifts to a level sufficiently low to enable the observation of an Earth-like exoplanet around a Sun-like star \citep[e.g.,][]{Lay:2004}. A solution proposed to overcome this problem is to use more than two telescopes and phase chopping, which consists in synthesizing two different transmission maps with the same telescope array, by applying different phase shifts in the beam combination process. In addition to allowing more precise differential measurements, it is also possible to isolate the planetary signal from the contributions of symmetric brightness emissions such as the star, local zodiacal cloud, exozodiacal cloud, stray light, thermal, or detector gain. After the investigation of several interferometer architectures \citep[e.g.,][]{Angel:1997b,Mennesson:1997}, two array architectures have been thoroughly investigated by ESA during two parallel assessment studies carried out by EADS Astrium and Alcatel-Alenia Space in 2005-2006: the four-telescope X-array and the Three-Telescope Nuller \citep[TTN,][]{Karlsson:2004}. These studies included the launch requirements, payload spacecraft, and the ground segment during which the actual mission science would be executed. Almost simultaneously, NASA/JPL initiated a similar study for TPF-I and focused in particular on the Dual-chopped Bracewell \citep{Lay:2004} and X-array configurations. These efforts on both sides of the Atlantic have finally resulted in a convergence and consensus on mission architecture, the so-called non-coplanar (aka Emma-type) X-array. The baseline design consisted in four collector spacecraft, flying in rectangular formation and feeding light to the beam combiner spacecraft located approximately 1200\,m from the array. This arrangement makes available baselines up to 170\,m for nulling measurements and up to 500\,m for the general astrophysics program (constructive imaging with an angular resolution of 4\,mas at 10~$\mu$m).\\


\subsection{Exoplanet yield}\label{sec:yield}

The number of planetary atmospheres that can be studied with such an instrument is a critical metric to estimate the science return of the mission. During the \darwin/TPF studies, the exoplanet yield has been extensively studied and cross-validated between ESA and NASA using various assumptions on the existing exoplanet population and prevalence of exozodiacal dust, which were both unknown at that time \citep[e.g.,][]{Defrere:2010}. Today, the occurrence of HZ rocky exoplanets has been measured by Kepler \citep[e.g.,][]{Winn:2015} and the prevalence of exozodiacal dust has been very well constrained by ground-based nulling interferometers \citep[][Ertel et al. in prep]{Mennesson:2014}. It is therefore possible to predict the exoplanet yield of direct imaging instruments more precisely. Recently, based on planet occurrence statistics from Kepler and using Monte-Carlo simulations, \citet[][]{Kammerer:2017,Quanz:2018} estimated that $\sim$315$_{-77}^{+113}$ exoplanets with radii between 0.5~R$_\oplus$ and 6~R$_\oplus$ can be detected during $\sim$0.52 years of mission time assuming four 2-m apertures and throughputs 3.5 times worse than those for the JWST and 40\% overheads. Approximately 85 planets could be habitable (radii between 0.5~R$_\oplus$ and 1.75~R$_\oplus$ and equilibrium temperatures between 200\,K and 450\,K) and would be prime targets for follow-up spectroscopic observations. While this exoplanet yield can be interpreted as an upper limit because no exozodiacal dust was assumed in this study, new results from the NASA's Hunt for Observable Signatures of Terrestrial planetary Systems \citep[HOSTS,][]{Danchi:2014} survey on the Large Binocular Telescope Interferometer \citep[LBTI,][]{Hinz:2016} suggest that exozodiacal dust disks would not be a major source of noise \citep{Ertel:2018a,Ertel:2018b}. New science yield estimates based on HOSTS upper limits are currently under study. 


\subsection{Technology state-of-the-art} 
\subsubsection{Formation flying} \label{sec:ff}

\begin{figure}[!t]
	\begin{center}
		\includegraphics[height=4.6 cm]{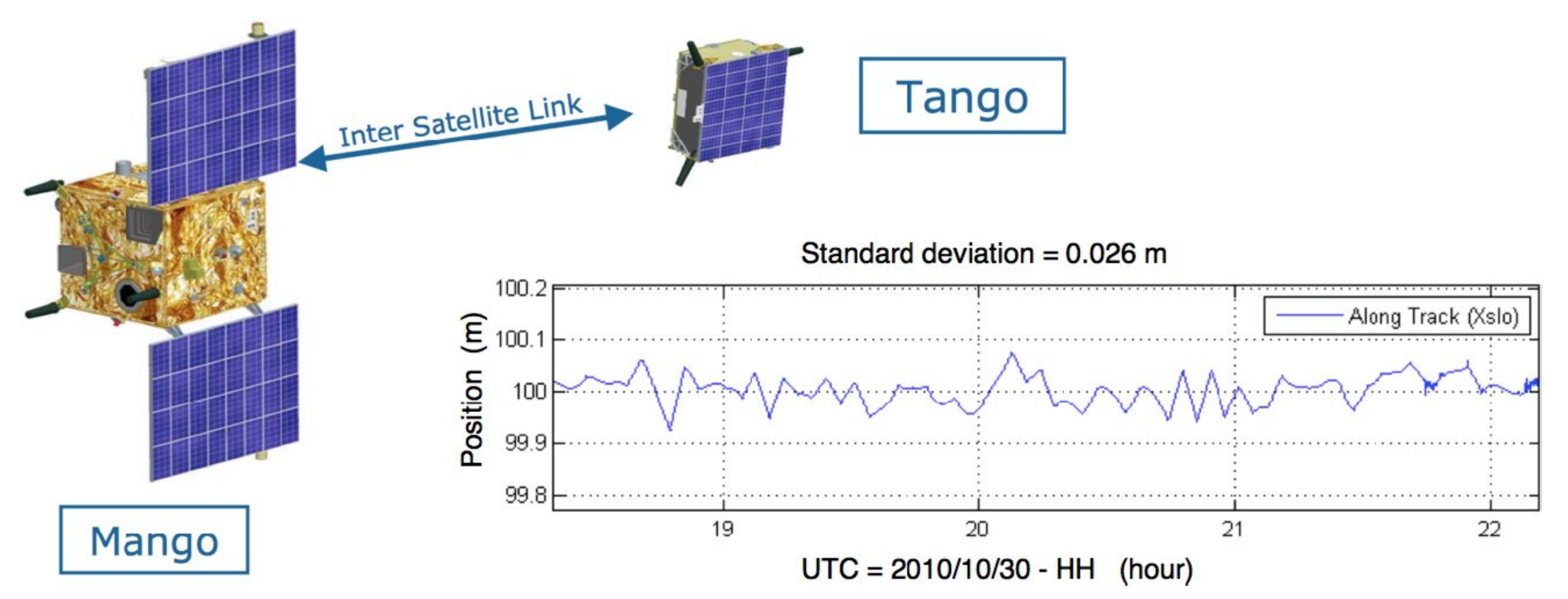}
		\caption{Demonstration of Formation Flying between the two PRISMA spacecraft, Tango and Mango. A distance of 100\,m was maintained during four hours with a standard deviation of few centimeters (see bottom right panel), limited by the accuracy of the radio frequency sensors. ESA's Proba 3 instrument, scheduled for a launch in 2021, should reduce this error to a few 100 microns.\textcopyright~Swedish Space Corporation, CNES and DLR.}\label{fig:prisma}
		\vspace{-1.5em}
	\end{center}
\end{figure}

Formation flying is a key technology for the deployment and success of a space-based interferometer. Remarkable advances in technology have been made in Europe in recent years with the space-based demonstration of this technology by the PRISMA mission \citep{Damico:2012}. PRISMA demonstrated a sub-cm positioning accuracy between two spacecraft (see Figure~\ref{fig:prisma}), mainly limited by the metrology system (GPS and RF). The launch of ESA's PROBA-3 mission in 2020 will provide further valuable free-flyer positioning accuracy results (sub-mm), which exceeds the requirements for a space-based interferometer that relies on fast pathlength correctors for precise Optical Path Delay (OPD) control. Extending the flight-tested building-block functionality from a distributed two-spacecraft instrument to an instrument with more spacecraft mainly relies on the replication of the coordination functionality and does not present additional complexity in terms of procedures according to the PRISMA navigation team. While formation flying can then be considered to have reached a technical readiness level (TRL) of 9 once PROBA-3 has flown, an uncertainty remains regarding fuel usage and the possible lifetime of such a mission. 

\subsubsection{Spatial filters}

Spatial filters are optical devices which significantly reduce the optical aberrations in wavefronts. Consequently, they are very important for nulling interferometry, making extremely deep nulls possible. To provide spatial filtering over large bandwidths, a variety of techniques can be applied, including single-mode fiber optics, photonic crystal fibers, or integrated optics. Developments of single-mode fibers for the mid-infrared were funded by NASA between 2003 and 2008 \citep[][]{Ksendzov:2007,Ksendzov:2008}. Fiber optics made of chalcogenide and silver halide materials have been demonstrated to yield 25 dB or more rejection of higher-order spatial modes at 10~$\mu$m, but they would require the division of the 6--20\,$\mu$m band into two parts. Although this performance is sufficient for flight, it would be greatly advantageous to improve the throughput of these devices and to test them over the full wavelength range in cryogenic conditions. Spatial filter technology would then be at TRL 5. The spatial filtering capabilities of photonic crystal fibers should also be investigated for use at mid-infrared wavelengths, because of the improved throughput that they may provide and the possibility to cover the whole wavelength range with a single device. Note finally that, given the recent developments in wavefront control with extreme adaptive optics systems \citep[e.g.,][]{Jovanovic:2015}, it is not clear whether such a filtering technology will be required. This should be addressed in the future.  

\subsubsection{Beam combination}

Classical optical designs of four-telescope nulling beam combiners \citep{Martin:2010} have been demonstrated to flight requirement levels albeit at room temperature and using signals much stronger than those from astronomical sources. Recently, a very promising grating nuller approach was proposed and shown to achieve nulls of 4$\times$10$^{-5}$  over the full 18\% bandwidth K-band \citep{Martin:2017}. Alternatively, {\it integrated optics} (IO) beam combination can be achieved by a network of single-mode waveguides embedded in a cm-scale glass chip. This solution bypasses complex optical interferometric trains sensitive to vibrational, thermal and mechanical stress, hence reducing the risk associated with bulky science instruments. As for fibers, IO can furthermore achieve wavefront cleaning to mitigate phase errors \citep{Wallner:2002,Mennesson:2002}. Silica-based IO solutions are now being successfully implemented in operating near-infrared 4-telescope interferometers \citep{Eisenhauer:2011,Lebouquin:2011} and have also been investigated for nulling applications in the near-infrared at 1.5\,$\mu$m \citep{Weber:2004,Errmann:2015} with stable nulls down to 10$^{-4}$ over 5\% bandwidth. Since silica glasses are opaque to IR radiation for $\lambda$$>$2\,$\mu$m, the extension of the IO approach to the mid-infrared in the 3-30\,$\mu$m range requires an adequate material and technological platform to manufacture high quality optical chips.\\ 

\subsubsection{Starlight suppression}\label{sec:star_suppr}

\begin{figure}[!t]
	\begin{center}
		\includegraphics[height=5.00 cm]{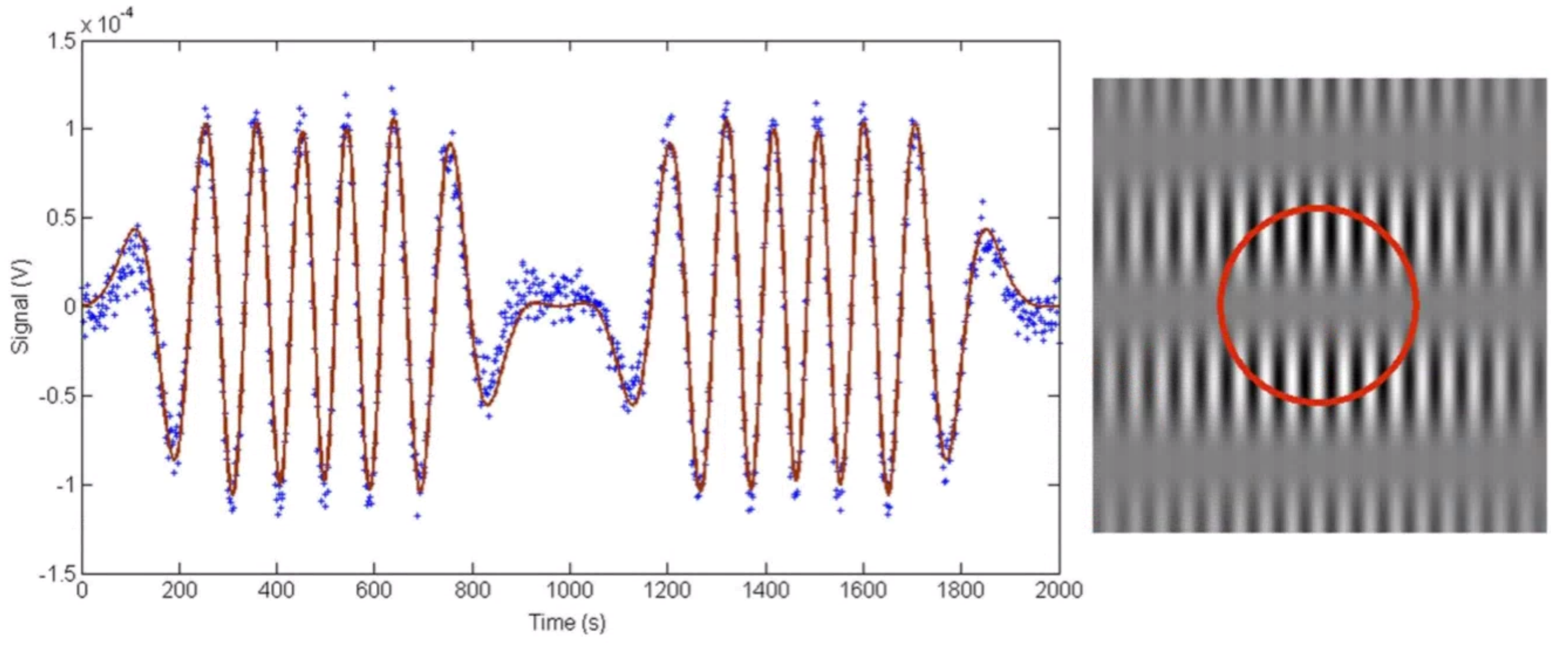}
		\caption{The left panel shows the planet signal detected with the Planet Detection Testbed \citep{Martin:2012}. Each point is an item of data from the 2-s chop cycle and the whole trace shows the null signal obtained over a 360 degree effective rotation of the interferometer array. The line is a fit to the signal from a planet at a nominal angular radius of $6.35\times10^{-7}$~rad (or 132\,mas) from the star. By comparison, the equivalent angular fringe distance from the short baseline is $4.7\times10^{-7}$~rad. Near the center and at the ends of the plot, the planet crosses the null fringe. The right panel shows the equivalent sensitivity map of the interferometer array. Array rotation causes the planet location to orbit (solid line) around the central null fringe (gray), and thus its signal is modulated both by the higher frequency fringes on the long baseline and by the chopping. {\bf Note that this demonstration was realized at room temperature with high flux levels. Technological developments are required to repeat this demonstration in cryogenic conditions with optical fluxes similar to those expected from astronomical sources (see Section~\ref{sec:tec})}.}\label{fig:nulling}
		\vspace{-1.5em}
	\end{center}
\end{figure}

A considerable expertise has been developed in the field of starlight suppression over the past 20 years, both in academic and industrial centers across the globe. These efforts culminated with laboratory demonstrations of the recombination scheme to flight requirements at the \emph{Jet Propulsion Laboratory} (JPL) in the US. In particular, mid-infrared nulls of 10$^{-5}$ were achieved with both the Adaptive Nuller \citep{Peters:2010} and the planet detection testbed \citep[][see Figure~\ref{fig:nulling}]{Martin:2012}, but with fluxes much higher than those expected from stars and planets allowing working at room temperature without being disturbed by the thermal emission of the environment. In parallel, the operation of high-precision ground-based interferometers has matured in both Europe and the US. In particular, Europe has gained a strong expertise in the field of fringe sensing, tracking, and stabilization with the operation of the \emph{Very Large Telescope Interferometer} (VLTI). In the United states, considerable technical expertise was gained by operating several nulling interferometers such as the Keck Interferometer Nuller \citep[KIN,][]{Colavita:2009}, the Palomar Fiber Nuller \citep[PFN,][]{Mennesson:2011a}, and the Large Binocular Telescope Interferometer \citep[][]{Hinz:2016}. All have produced excellent scientific results \citep[e.g.,][]{Mennesson:2014,Defrere:2015}, which have pushed high-resolution mid-infrared imaging to new limits \citep{Defrere:2016}. Innovative data reduction techniques have also been developed to improve the accuracy of nulling instruments \citep{Hanot:2011,Mennesson:2011b} but more work is required to adapt this technique to four-telescope configurations.


\section{Prospects}

\subsection{Current context}

\begin{figure}[!t]
	\begin{center}
		\includegraphics[height=7.8 cm]{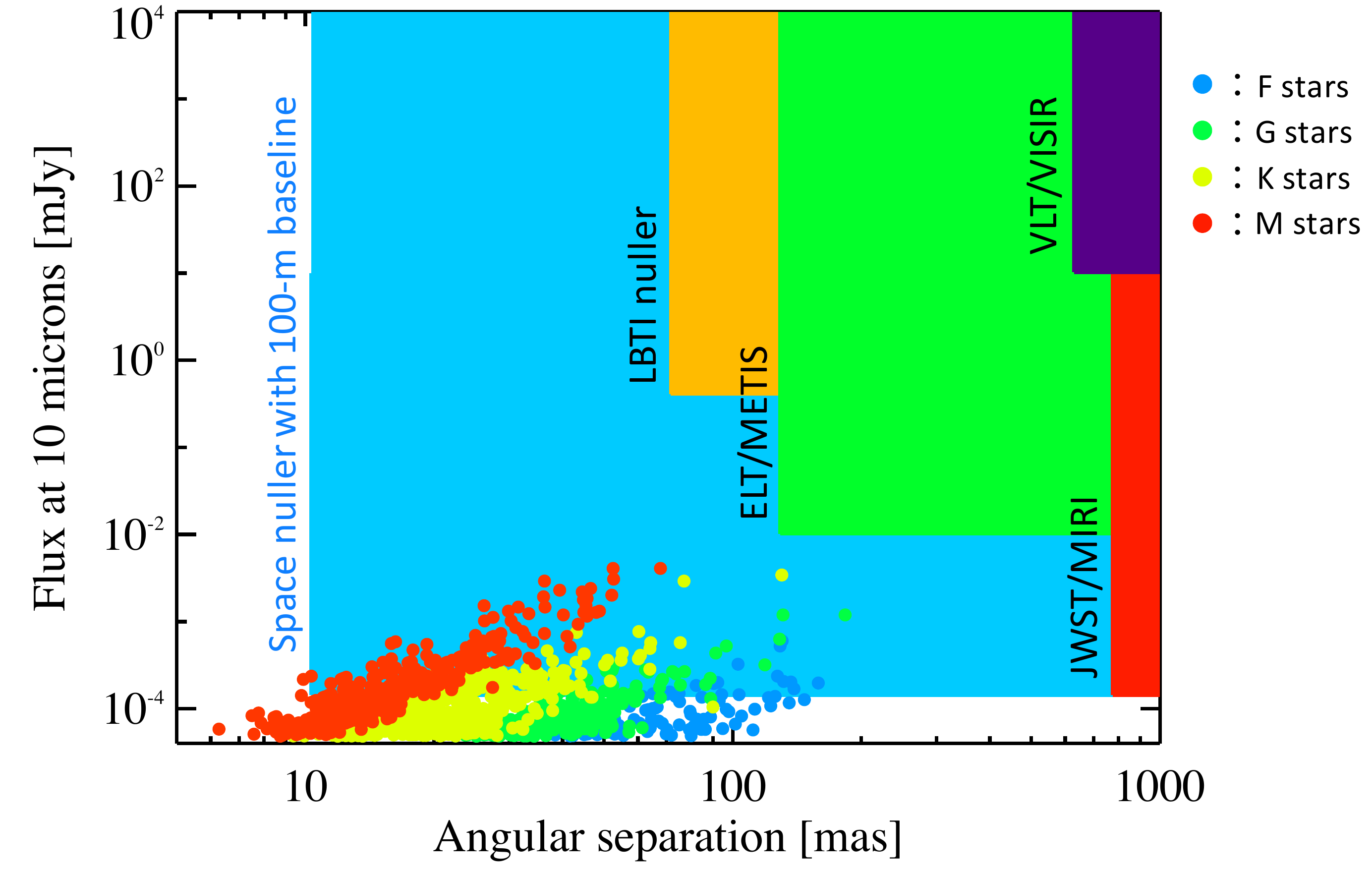}
		\caption{Flux at 10~$\mu$m as a function of angular separation of putative 300\,K blackbody planets located in the middle of the HZ of nearby single main-sequence stars \citep[DASSC catalog,][]{Kaltenegger:2010}. The 10-$\sigma$ sensitivity in one day of current and planned mid-infrared instruments is indicated by the colored rectangles \citep[values scaled from][]{Brandl:2014} and compared to that of a four 2m-aperture space-based nulling instrument (see blue rectangle, assuming no exozodiacal dust). Note that this plot does not take into account aperture masking modes, which improve the angular resolution but reduce the sensitivity.}\label{fig:synergies}
		\vspace{-1.5em}
	\end{center}
\end{figure}

Because it is a very challenging observational task partially requiring further technological developments, the spectroscopic characterisation of small HZ exoplanets by direct imaging is currently not possible with existing instruments. Future instruments considered to achieve this goal cover generally two distinct and complementary wavelength ranges: the visible, which favors the angular resolution, and the mid-infrared, which favors the contrast. Focusing first on the mid-infrared regime, we can see in Figure~\ref{fig:synergies} that none of current or foreseen instruments can approach a space-based nulling interferometer when it comes to achieving the necessary sensitivity at a small angular separation from the parent star. In the case of JWST, the impressive sensitivity provided by the large collecting area (25~m$^2$) and cold (40\,K) telescope optics can only be utilized by coronagraphs which are expected to achieve a best case contrast at 10.6~$\mu$m of 10$^{-4}$ to 10$^{-5}$, for separations larger than 0.5 to 1.0 arcsecond \citep[][]{Boccaletti:2015}. With such performances, the detection of warm and young exo-Jupiters is the closest that the JWST/MIRI instrument can approach to this project's goal of exo-Earth characterisation. For the E-ELT, the massive gain in collecting area (980~m$^2$) compared to the JWST offsets the impact of having warm optics to give comparable sensitivity limits for the METIS instrument, operating at 3 to 19~$\mu$m. METIS \citep[e.g.,][]{Brandl:2016} will be equipped with coronagraphs which can in principle achieve contrasts of $\sim$10$^{-7}$ at separation of $\sim$0.7 arcsecond necessary to directly image a putative exo-Earth orbiting $\alpha$~Cen and $\sim$10 small planets (1 to 4 R$_\oplus$) with equilibrium temperatures between 200 and 500\,K around the nearest stars \citep{Quanz15}. However, achieving this performance in practice at a ground-based observatory where image quality and stability are dependent on an advanced adaptive optics system, will be challenging. Further, due to the scarcity of available photons, the measurement would be restricted to a photometric detection, with little hope of spectroscopic follow-up.\\ 

Regarding the characterisation of HZ exoplanets in the visible, NASA is currently studying two concepts in preparation for the 2020 US decadal survey in Astronomy: (i) LUVOIR, a 10-16 m segmented, visible light telescope designed for an ambitious program of general astrophysics as well as detection of Earth-sized exoplanets and characterisation of dozens to hundreds of nearby stars; and (ii) HabEx, a 4-8 m monolithic telescope optimized for detection of Earth-sized exoplanets and characterisation of a smaller number of systems using either a highly optimized coronagraph or possibly a star shade. Both missions will likely include spectroscopy in the visible to near-infrared of Earth-sized planets in the HZ of nearby stars, searching for signs of habitability (H$_{2}$O) and bio-signature gases (O$_{2}$, O$_{3}$). A possible near-infrared extension (up to $\sim$2.5 microns) of these high contrast spectroscopic capabilities would help further establish whether these gases were created by biotic processes or not, i.e., looking for species such as CO$_{2}$, CO, O$_{3}$, CH$_{4}$, and N$_2$O. {\bf Scientifically, there is no clear advantage of one wavelength range over the other as described in various reviews \citep[see e.g.,][]{Desmarais:2002,Fujii:2018}. Note that mid-infrared observations may also be used to estimate the planetary radius, which otherwise remains unobservable unless the planet transits.}

\subsection{Precursor concepts}

The path towards space-based interferometry is regularly discussed and this often involves the need for precursor missions \citep[e.g.,][]{Rinehart:2016}. In addition to free-flying demonstrators already discussed in Section~\ref{sec:ff},  concepts of small-scale space-based infrared nulling interferometers were seriously considered both in Europe and in the US: the Fourier-Kelvin Stellar Interferometer \citep[FKSI,][]{Danchi:2008} and Pegase \citep[][]{Ollivier:2009}. Opportunities for testing technologies and for pushing new developments also exist with current ground-based facilities. For instance, in addition to ongoing research with current instruments as discussed in Section~\ref{sec:star_suppr}, there is currently a plan to build a nulling interferometer for the VLTI \citep[i.e., the Hi-5 project,][]{Defrere:2018b}. There is also a science-driven, international initiative to develop the roadmap for a future ground-based facility that will be optimised to image planet-forming disks on the spatial scale where the protoplanets are assembled, which is the Hill Sphere of the forming planets. This \textit{Planet Formation Imager} \citep[PFI,][]{Monnier:2016} is designed to detect and characterise protoplanets during their first $\sim 100$ million years and trace how the planetary population changes due to migration processes, unveiling the processes that determine the final architecture of exoplanetary systems. With $\sim 20$ telescope elements and baselines of $\sim 3$~km, the PFI concept is optimised for imaging complex scenes at mid-infrared wavelengths ($3-12\mu$m) and at 0.1 milliarcsecond resolution. This clearly complements the capabilities of a space interferometer that would be optimised to achieve the sensitivity and contrast required to characterise the atmospheres of mature exoplanets. 

\subsection{Required technological developments}\label{sec:tec}

The main remaining technological challenge is the implementation of a cryogenic interferometer system that achieves the necessary starlight suppression and actual planet detection from 5 to 20~$\mu$m with optical fluxes similar to those expected from astronomical sources (typically $\sim$0.2 photons/s/m$^2$ for an Earth-like planet located at 10\,pc). To achieve this goal, preliminary system studies are required to (i) define the cryogenic design for passive cooling of the optics and active cooling of the detectors; (ii) characterize and minimize the vibrations of the interferometer in cryogenic conditions; (iii) validate the cryogenic deformable mirrors, and (iv) develop spatial filters (if needed) and beam combiners that can provide the necessary performance from 6 to 20~$\mu$m under cryogenic conditions. Specific developments in terms of fringe tracking (taking into account residual vibrations) and data reduction will undoubtedly be needed to reach the required level of performance in terms of starlight suppression. Dedicated developments will also be required in the field of mid-infrared detectors, although the JWST legacy will be particularly useful in this context.

\section{Conclusions}

Mid-infrared space-based interferometry is a technology of direct imaging that can uniquely characterise the atmospheres of terrestrial exoplanets around nearby main-sequence stars. The use of formation-flying telescopes makes it possible to observe and study a wide variety of planetary systems, including HZ terrestrial planets around M dwarfs such as Proxima b \citep{Defrere:2018}. Currently, no other technology can obtain mid-infrared spectra of a statistically-meaningful number of temperate rocky exoplanets, which is required to make progress towards the search for life in the Universe. Significant investments would have to be made today in order to ensure that the development of such an instrument is possible in the not too distant future.  
\section*{Acknowledgements}

This work was partly funded by the European Research Council under the European Union's Seventh Framework Program (ERC Grant Agreement n.~337569) and by the French Community of Belgium through an ARC grant for Concerted Research Action. DD and OA acknowledge funding from the FRS-FNRS. Some of research described in this publication was carried out in part at the Jet Propulsion Laboratory, California Institute of Technology, under a contract with the National Aeronautics and Space Administration. Part of this work has been carried out within the frame of the National Center for Competence in Research PlanetS supported by the Swiss National Science Foundation. SPQ acknowledges the financial support of the SNSF. NCS was supported by Funda\c{c}\~ao para a Ci\^encia e a Tecnologia (FCT, Portugal) through national funds and by FEDER through COMPETE2020 in the context of the projects and grants reference UID/FIS/04434/2013 \& POCI-01-0145-FEDER-007672, PTDC/FIS-AST/1526/2014 \& POCI-01-0145-FEDER-016886, and IF/00169/2012/CP0150/CT0002. SL acknowledges support from ERC Starting Grant n.~639248. SK acknowledges support from an STFC Rutherford Fellowship (ST/J004030/1) and ERC Starting Grant n.~639889.

\bibliographystyle{spbasic}        

\end{document}